\documentclass[prb,twocolumn]{revtex4}
\usepackage{graphicx}
\usepackage{dcolumn}
\usepackage{bm}
\usepackage{amsmath}
\usepackage{amsfonts}
\usepackage{bbm} 
 
\newcommand{\bra}[1]{\langle #1|}
\newcommand{\ket}[1]{|#1\rangle}

\begin{document}

\title{Ultrastrong coupling between a cavity resonator \\
 and the cyclotron transition of a 2D electron gas in the case of integer filling factor}
  
\author{David Hagenm\"{u}ller$^1$}
\author{Simone De Liberato$^{1,2}$}
\author{Cristiano Ciuti$^1$}
\email{cristiano.ciuti@univ-paris-diderot.fr}
\affiliation{$^1$ Laboratoire Mat\'eriaux et Ph\'enom\`enes Quantiques,
Universit\'e Paris Diderot-Paris 7 and CNRS, \\ B\^atiment Condorcet, 10 rue
Alice Domont et L\'eonie Duquet, 75205 Paris Cedex 13, France}
\affiliation{$^2$ Department of Physics, University of Tokyo, Hongo, Bunkyo-Ku, Tokyo 113-8656, Japan}
 
\begin{abstract}
We investigate theoretically the coupling between a cavity resonator and the cyclotron transition of a two dimensional electron gas under an applied perpendicular magnetic field. We derive and diagonalize an effective quantum Hamiltonian describing the magnetopolariton excitations of the two dimensional electron gas for the case of integer filling factors. The limits of validity of the present approach are critically discussed. The dimensionless vacuum Rabi frequency $\Omega_0/\omega_0$ (i.e., normalized to the cyclotron frequency $\omega_0$) is shown to scale as $\sqrt{\alpha\: n_{QW} \nu}$, where $\alpha$ is the fine structure constant, $n_{QW}$ is the number of quantum wells and $\nu$ is the filling factor in each well. We show that with realistic parameters of a high-mobility semiconductor two dimensional electron gas, the dimensionless coupling  $\Omega_0/\omega_0$ can be much larger than 1 in the case of $\nu \gg 1$, the latter condition being typically realized for cyclotron transitions in the microwave range. Implications of such ultrastrong coupling regime  are discussed. 
\end{abstract}

\maketitle
 
The study of light-matter coupling in confined geometries has been in the last two decades a very interesting topic in atomic and condensed matter physics. The realization of extremely high quality mirrors and the ability to manipulate atomic states with extremely long lifetimes has made possible the observation of the strong light-matter coupling regime. In this regime, the coupling exceeds the losses and it is therefore possible to spectroscopically resolve the so-called vacuum Rabi splitting\cite{Thompson92,Raymond01}. Strong light-matter coupling has been observed in various solid state systems, ranging from microcavity embedded quantum wells \cite{Weisbuch92,Dini03} to superconducting circuits coupled to transmission line resonators \cite{Wallraff04}. 
In solid state systems, thanks to the collective nature of the excitations, it is possible to go further the standard strong coupling regime and to enter a new regime, where the vacuum Rabi frequency is not only larger than the loss rate, but becomes comparable or larger than the bare frequency of the uncoupled excitations \cite{Ciuti05, Ciuti06, Devoret07,Anappara09,Yanko09,Yanko10,Bourassa,Nataf}. 
The ultrastrong coupling regime is interesting for example for non-adiabatic cavity QED phenomena reminiscent of the dynamical Casimir effect\cite{Kardar99,DeLiberato07,Gunter09,DeLiberato09}. Moreover,  it can lead to a dramatic modification of the quantum ground state (vacuum) properties\cite{Ciuti05,Nataf}. 

In this paper, we investigate the coupling of the magnetic cyclotron transition of a two-dimensional electron gas (2DEG) to the quantum field of a cavity resonator. We show that the dimensionless vacuum Rabi frequency $\Omega_R/\omega_0$  can be largely enhanced with respect to the case of intersubband transitions in a 2DEG without magnetic field. In particular, this is the case in the regime of high filling factors, obtained with relatively weak magnetic fields and large electron densities, which can be obtained in state-of-the-art high mobility 2DEGs. We derive the second quantized Hamiltonian for such a system in the case of integer filling factor $\nu$ and derive an effective Hamiltonian describing the cavity magnetopolariton excitations of the 2DEG. We show that since $\Omega_R/\omega_0 \gg 1$, the diamagnetic ${\bm A}^{2}$-term of the quantum light-matter coupling Hamiltonian becomes dominant in such a system. The present work is relevant not only for the fundamental quantum electrodynamical properties of a 2DEG in a unconventional regime. It may have important implications also in the low-frequency magnetotransport properties of the 2DEG embedded in a cavity resonator. 

The paper is structured as follows. In Sec. \ref{system}, we introduce in detail the system and show why the magnetic cyclotron transition can be ultrastrongly coupled to the vacuum field of a cavity. In Sec. \ref{Hamiltonian}, we present the second quantized quantum light-matter Hamiltonian for the system (details about the derivation are given in the Appendix). In Sec. \ref{Results}, we diagonalize an effective bosonic Hamiltonian to describe the magnetopolariton excitations of the 2DEG and show the resulting mode dispersions. Conclusions and perspectives are drawn in Sec. \ref{Conclusions}.

\section{Physical system and scaling of coupling}
\label{system}
\begin{figure}[t!]
\begin{center}
\includegraphics[width=200pt]{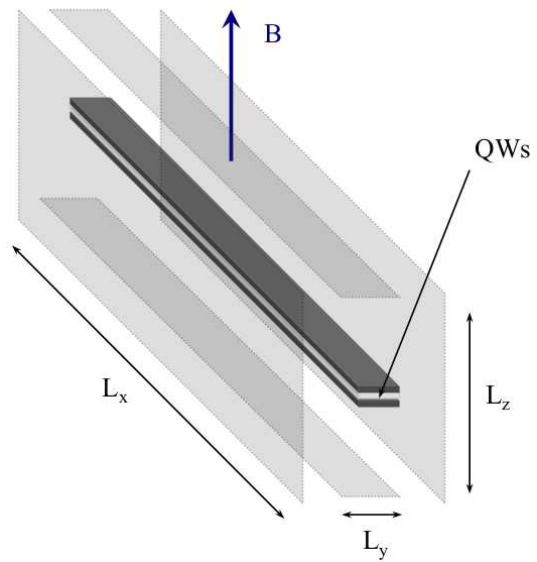}
\caption{\label{cavite} \it Sketch of a cavity resonator embedding $n_{QW}$ identical quantum wells (QWs), each containing a two-dimensional electron gas (2DEG) in the $xy$ plane.  An uniform and static magnetic field $B$ is applied along the $z$ axis. } 
\end{center}
\end{figure}

We will consider a system consisting of multiple doped semiconductor quantum wells (QWs) in presence of a magnetic field $B$ along the $z$ axis (perpendicular to the QW plane). The QWs are embedded in a wire-like cavity resonator, as depicted in Fig. \ref{cavite}, that confines the electromagnetic modes along two directions ($z$ and $y$).

\begin{figure}[t!]
\begin{center}
\includegraphics[width=270pt]{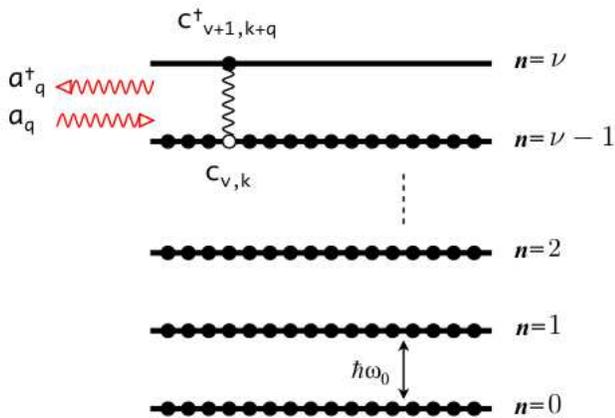}
\caption{\it \label{Landau} Sketch of Landau levels with an integer filling factor $\nu$ ($\nu \gg 1$ is the regime considered in this paper). The cyclotron transition between levels $\nu-1$ and $\nu$  is resonantly coupled to cavity mode quantum field.}
\end{center}
\end{figure}

In presence of a magnetic field, the electrons occupy highly degenerate bands (see Fig. \ref{Landau}), the well-known Landau levels (LLs), separated by the cyclotron energy equal to $\hbar \omega_{0}$, where $\omega_{0}=e B/(m^{*}c)$ is the cyclotron frequency ($m^{*}\approx 0.068 \,m_{0}$ is the effective electron mass of the conduction band in GaAs). The magnetic length, associated to the fundamental Landau level, is $l_{0}=\sqrt{\hbar/(m^{*}\omega_{0})}$, while the degeneracy of each LL (taking into account for the electron spin) is  $\mathcal{N}= S/(\pi l^{2}_{0})$, where $S$ is the surface of the sample. 

Here, we will consider the case of an integer filling factor $\nu$, i.e., electrons fill completely the first $\nu$ LLs (the LL index goes from $n=0$ to $n=\nu-1$). Consequently, $\nu =\rho_{2DEG} S/\mathcal{N}$ where $\rho_{2DEG}$ is the electron density per unit of area in each QW. In this section, we wish to show a back-of-the-envelope calculation of the vacuum Rabi frequency for such a system and compare it to the bare cyclotron transition frequency $\omega_0$. This simple calculation will allow the reader to grasp the essential quantitative features and key parameters. A complete and rigorous derivation will be presented in Sec. \ref{Hamiltonian} and in the Appendix. As we are interested in the observation of high filling factor LLs, we have to consider a sample at cryogenic temperature. In the rest of the manuscript we will thus perform calculations at $T=0$, even if our results are expected to be robust while the thermal energy does not exceed the cyclotron transition energy ($k_BT<\hbar \omega_0$). 

Due to Pauli blocking and harmonic oscillator selection rules, only electrons populating the level $n=\nu-1$ will participate to the light-matter coupling. Given that the typical cyclotron radius for an electron in  
%LL $n_{0}$ with kinetic energy $n_{0} \hbar \omega_{0}$ is $r_{0} \sim l_{0} \sqrt{n_{0}}$, the electric dipole associated to the transition from level $n=\nu-1$ to level $n=\nu$ is  $d \sim e \, l_0 \sqrt{\nu}$.
 LL $n$ is $r \sim l_{0} \sqrt{n}$, the electric dipole associated to the transition from level $n=\nu-1$ to level $n=\nu$ will be  $d \sim e \, l_0 \sqrt{\nu}$.
Hence, larger filling factors produce a larger dipole (in a way reminiscent to Rydberg atomic states with large orbital principal quantum number). 
However, if $N_{2DEG} = \rho_{2DEG} S $ is the number of electrons in the 2DEG, only the number $N_{2DEG}/\nu$ in the level $n=\nu -1$ is optically active.
It is known that in the presence of a collection of identical dipoles, the collective vacuum Rabi frequency is proportional to the square root of the number of dipoles \cite{Dicke54}. Hence, the collective excitation of the 2DEG enhances the light-matter coupling by a factor $\sqrt{N_{2DEG}/\nu}$. If $n_{QW}$ quantum wells are identically coupled to the field, an additional enhancement of  $\sqrt{n_{QW}}$ can be obtained (we consider here that the QWs are not electronically coupled; this is therefore equivalent to simply increase the density of the 2DEG).

The vacuum Rabi frequency is therefore $\hbar \Omega_{res} =  d \times E_{vac} \times \sqrt{N_{2DEG} \, n_{QW} /\nu}$, where $E_{vac}$ is the electric field associated to the cavity vacuum fluctuations.  We will call $V = L_z S$ the volume of the cavity, with  $L_z$ being the direction orthonormal to the 2DEG and $S=L_{x} L_{y}$ the surface of the 2DEG plane (cf figure \ref{cavite}). Now, if we consider a cavity mode resonant with the cyclotron transition, we have $E_{vac} \sim \sqrt{\frac{\hbar \omega_0}{\epsilon L_z S}}$, where $\epsilon$ is the dielectric constant of the cavity spacer ( $\epsilon \simeq 13$ for GaAs structures). Finally, we find

\begin{equation}
\frac{\Omega_{res}}{\omega_0} \sim e \, l_0 \sqrt{\nu} \; \sqrt{\frac{\hbar }{\omega_0 \epsilon L_z S}} \; \sqrt{N_{2DEG} \, n_{QW} /\nu}. 
\end{equation}
If we consider a half-wavelength cavity, i.e., $L_z = \lambda_0/2$ and $\omega_0 = \frac{2 \pi}{\lambda_0} \frac{c}{\sqrt{\epsilon}}$, given that
 $\rho_{2DEG}\, \pi l_0^2 = \nu$, we obtain
\begin{equation}
\label{ratio}
\frac{\Omega_{res}}{\omega_0} \sim \sqrt{\alpha \, n_{QW} \nu},
\end{equation}
where $\alpha = \frac{e^2}{\hbar c} \simeq \frac{1}{137}$ is the fine structure constant.
Hence, the dimensionless vacuum Rabi coupling depends on the small fundamental constant $\alpha$, on the number of QWs and on the filling factor in each QW. For very large filling factors and a large number of QWs it is thus possible to have $\frac{\Omega_{res}}{\omega_0}  \gg 1$. For a given density of electrons, since $\nu \propto 1/\sqrt{B}$, we can thus always increase the coupling by lowering the magnetic field intensity, since for $B \to 0$, $\nu \to  + \infty$. One the other hand, for a given magnetic field, the filling factor increases with increasing the electron density, and so does the coupling strength. Of course, this description makes sense only if the cyclotron resonance is well resolved (i.e., not quenched by the broadening). For example, as shown by the experimental work in Ref. \onlinecite{Ritchie}, in GaAs 2DEG with a relatively high mobility ($\mu = 1.6 \; 10^{6}\; {\rm cm^{2}V^{-1} s^{-1}}$), it is possible to have at the same time very high filling factors ($\nu > 100$) with a well resolved cyclotron resonance for a magnetic field $B =0.01$ T and a transition frequency in the microwave range ($f_0 = \frac{\omega_0}{2 \pi} \approx 30$ GHz).

\vspace{3mm}

\section{Quantum Hamiltonian}
\label{Hamiltonian}

\subsection{General considerations}
\label{subA}
In condensed matter systems consisting of a collection of two-level systems, in the low excitation limit,  it is generally possible to describe the collective excitations as bosons. Starting from the Hamiltonian in terms of the electronic fermion operators and of the photonic boson operators (see Appendix), we have derived the effective bosonic Hamiltonian describing the magnetopolariton excitations.  In this case, the Hamiltonian is exactly solvable by mean of an Hopfield-Bogoliubov diagonalization. This procedure has been already applied to other systems: for example, exciton polaritons in bulk materials\cite{Hopfield} and quantum wells\cite{Savona94,Gerace07}, intersubband polaritons \cite{Ciuti05} and polariton in atomic gases \cite{Carusotto08}.
Among the different possible geometries,  we chose to treat here the case of a photonic asymmetric wire ($L_x \gg L_z \gg L_y$). This will permit us to treat the continuous dispersion along the $x$ axis, while keeping only few discrete modes along the $z$ axis and neglecting completely all the modes along the $y$ axis except the lowest one, whose intensity profile is constant over the sample surface. 
While this choice is motivated both by the simplicity of the resulting Hamiltonian and by considerations of experimental feasibility (using transmission line resonators) the extension to other one dimensional geometries does not present any particular difficulty.
For sake of simplicity, we will consider the quantum wells to be placed in the middle of the photonic wire, at $z=L_z/2$.

The reader familiar with the literature on polaritons will notice that here, even in the resonant case (one photonic mode resonant with the cyclotron frequency) we will not neglect the higher lying photonic modes, as often done in the literature. We will instead formally consider all of them, cutting then the resulting infinite matrix in order to retain enough modes to be at convergence. While considering only one photonic mode remarkably simplifies the algebra, it is not a valid approximation in the system under consideration. The spacing between photonic branches being constant, in the resonant case, the condition $\Omega_{0}/\omega_{0} > 1$ implies that also the higher photonic modes are coupled to the transition. 
\subsection{Coulomb interactions}

\label{coulomb}

As well known in the context of the fractional quantum Hall effect, the Coulomb interaction can play a crucial role in two-dimensional systems of electrons under magnetic field. It is now well understood that the role of interactions is completely different depending on whether one considers the case of integer or fractional filling factor. In the case of fractional filling, correlation can become crucial, because a rearrangement of a many-electron configuration within a partially filled Landau band does not cost any kinetic energy. As it has been said in the previous sections, we consider here only the case of an integer filling factor $\nu$ in the peculiar weak magnetic field regime such that $\nu \gg 1$. In this weak magnetic field limit, we have $e^{2}/ (\epsilon \l_{0}) \gg \hbar \omega_{0}$: hence, one might \textit{a priori} expect a significant impact of Coulomb interactions, which could produce for example a mixing of the Landau levels. However, it has been proved \cite{Aleiner} that screening leads to a renormalized Coulomb potential which abruptly drops at the distance around two cyclotron radii. Moreover, the effective interaction is  much smaller than $\hbar \omega_{0}$ and this allows us to treat only the electrons belonging to level $n=\nu-1$ (which is the only one optically active). In addition, Kohn's theorem \cite{Kohn13} states that the cyclotron resonance is not affected at all by electron-electron interactions as far as we consider a translationally invariant system. This holds in the absence of disorder and for a photonic wavevector $q=0$.
In the geometry we consider (cf sec. \ref{subA} and \ref{subB}), the photonic  wave vector always satisfies the condition that $q l_{0} \ll 1$ (even far from resonance). Indeed, since for $\nu \gg 1$ the vacuum Rabi frequency $\Omega_0$ can be even much larger than the cyclotron frequency $\omega_0$, the light-matter interaction in the cavity system  appears to be by far the most dominant interaction.

\subsection{Cavity quantized electromagnetic field}
\label{subB}
In the considered geometry, the vector potential can be written as 
\begin{equation}
\label{expressionA}
{\bm A} ({\bm r}) = {\bm A}_{0}({\bm r}) + {\bm A}_{em} ({\bm r}),
\end{equation}
where ${\bm A}_{0} ({\bm r})$ is the applied uniform magnetic field directed along the $z$-direction (${\bm A}_{0}=-By\, {\bm u}_{x}$ in the Landau gauge that we will use in this manuscript) and  ${\bm A}_{em}$ is the contribution associated to the cavity quantum field. We introduce the photon wavevector :
\begin{equation}
\label{mode2}
{\bm q} = \left( \begin{array}{c}   q_{x} \\ q_{y} \\ q_{z} \end{array} \right) = \left( \begin{array}{c}  q_x\\ \frac{\pi n_{y}}{L_{y}} \\ \frac{\pi n_{z}}{L_{z}} \end{array} \right),
\end{equation}
where $q_x$ can vary continuously, while $q_y$ and $q_z$ are quantized ($n_y$ and $n_z$ are integer values). 

In the following, we will take the length $L_z$ such that the mode corresponding to $q_x=0$, $q_y=0$ and $q_z=\frac{\pi}{L_z}$ is close to resonance with the cyclotron transition. 
As explained in Section \ref{subA},  the condition $L_y<<L_z$ allows us to neglect all the modes with $q_y \neq 0$. To simplify the notation, we will thus omit the $q_y=0$ index. 
The electromagnetic vector potential is then written as :

\begin{equation}
\label{vector1}
{\bm A_{em}}({\bm r}) =\sum_{q_x,n_z} \sqrt{\frac{2\pi\hbar c^{2}}{\epsilon \, \omega_{q_x,n_z}}} \left(a_{q_x,n_z} {\bm {\mathcal U}}_{q_x,n_z} + a^{\dagger}_{q_x,n_z} {\bm {\mathcal U}}^{*}_{q_x,n_z} \right),
\end{equation}

where the operator $a_{q_x,n_z}$ is the bosonic annihilation operator for a photon belonging to the mode labeled by the wavevector $\{q_x,q_z=\frac{\pi n_z}{L_{z}}\}$. The spatial shape ${\bm {\mathcal U}}_{q_x,n_z}$ of the modes (the derivation is presented in Appendix \ref{modes}) is given by : 

\begin{equation}
\label{shape}
{\bm {\mathcal U}}_{q_x,n_z} = \sqrt{\frac{2}{V}} e^{i q_{x} x} {\left( 
\begin{array}{c}
0 \\ \\ \sin (\frac{\pi n_z}{L_z} z) \\ \\ 0 \end{array} \right)} \\, 
\end{equation} 

Being our quantum wells positioned in the middle of the photonic cavity, at $z=L_z/2$, only the odd photonic modes will be coupled to the electron gas. Moreover we introduce the matrix notation :

\begin{equation}
{\bm a}_{q_x} \equiv \left(a_{q_x,1}\;a_{q_x,2} \;\cdots \;a_{q_x,n_z}\;\cdots\right)^{T}. 
\end{equation}

${\bm a}_{q_x}$ is thus a vector containing the photon annihilation operators for all the modes with different values of $q_z=\frac{\pi n_z}{L_z}$. 

\subsection{Light-matter coupled system}
\label{subC}
In the Landau gauge each electron state is indexed by three quantum numbers $n$, $k$ and $m$ (see Appendix \ref{appLandau} for details). $n$ indexes the different Landau levels, $k$ is the momentum component along $x$ and labels the different electrons in each Landau level. $m$ indexes the different subbands in the quantum well. Actually we show in Appendix \ref{appHamiltonian} that given the condition $L_{QW} \ll L_{z}$, we have the selection rule $m=m'$, meaning that we can safely neglect intersubband transitions and thus drop the $m$ index. Beside, one could argue that all the energies in play are much smaller than the intersubband gap which reinforces this approximation.

In order to write the second quantized Hamiltonian for the coupled light-matter system, we will thus introduce the fermionic operator $c_{n,k}^{(j)\dagger}$, that creates an electron with quantum number $k$ in the $n^{th}$ Landau level in the $j^{th}$ quantum well. We omit the spin quantum number because the optical transition conserves the electron spin, hence both spin channels contribute equally to the light-matter coupling: it will simply appears as an added degeneracy.  We also neglect the Zeeman splitting of the LLs (the cyclotron transition frequency is the same for both spin channels).

Analogously to the case of intersubband transitions\cite{Ciuti05}, the bright mode creation operator associated to the cyclotron transition is given by
\begin{equation}
b_{q_x}^{\dagger} = \sqrt{\frac{\nu}{n_{QW}\rho_{2DEG}S}}\; \sum_{j,k} c_{\nu,k+q_x}^{(j)\dagger} \, c_{\nu-1,k}^{(j)}.
\end{equation} 
%The prefactor is a normalization constant chosen such that $\langle 0 \vert b_{q_x} b^{\dagger}_{q_x} \vert 0 \rangle = 1$. In the dilute regime under consideration, these operators are approximately bosonic,  $[b_{q_x},b_{q'_x}^{\dagger}] \simeq \delta_{q_x,q'_x}$. The bright mode is  the collective excitation of the two dimensional electron gas which is directly coupled to the cavity photon mode.
The prefactor is a normalization constant chosen such that, in the dilute regime under consideration, these operators are approximately bosonic,  $[b_{q_x},b_{q'_x}^{\dagger}] \simeq \delta_{q_x,q'_x}$. The bright mode is  the collective excitation of the two dimensional electron gas which is directly coupled to the cavity photon mode.

After some calculations that are detailed in Appendix \ref{appHamiltonian}, we get the following Hamiltonian

\begin{equation}
H =  H_{Landau} + H_{int} + H_{dia} + H_{cavity},
\end{equation}

where

\begin{eqnarray}
\label{terms}
H_{Landau} &=&\sum_{q_{x}} \hbar \omega_{0} \, b_{q_x}^{\dagger}b_{q_x}, \nonumber \\  \nonumber \\ \\ 
H_{int} &=& \sum_{q_x} i \hbar \,{\bm \Omega}^{T}_{q_x} \; {\bm a}_{q_x} \left( b^{\dagger}_{q_x} -  b_{-q_x}  \right )\nonumber \\  \nonumber\\
&+& \sum_{q_x} i \hbar \,{\bm \Omega}^{T}_{q_x}\; {\bm a}^{\dagger}_{q_x}  \left(b_{-q_x}^{\dagger} - b_{q_x}\right ),\nonumber \\  \nonumber\\ \nonumber \\
H_{dia}&=&  \sum_{q_x} {\bm a}^{T}_{q_x} {\bm D}_{q_x} {\bm a}_{-q_x} + {\bm a}^{T}_{q_x}  {\bm D}_{q_x} {\bm a}^{\dagger}_{q_x} \nonumber \\  \nonumber\\
&+& {\bm a}^{\dagger T}_{q_x}  {\bm D}_{q_x} {\bm a}_{q_x} + {\bm a}_{q_x}^{\dagger T} {\bm D}_{q_x} {\bm a}_{-q_x}^{\dagger}, \nonumber
\end{eqnarray}

and

\begin{equation}
\label{field}
H_{cavity}  = \sum_{q_x} {\bm a}^{\dagger T}_{q_x} \hbar {\bm \omega}_{q_x} {\bm a}_{q_x},
\end{equation}

which represents the energy of the cavity quantum electromagnetic field. Note that we have omitted the zero-point energy that does not play any role here and we have defined 
\begin{equation}
{\bm \omega}_{q_x} = diag\left[\omega_{q_x,1}\;\omega_{q_x,2} \;\cdots \;\omega_{q_x,n_z}\;\cdots\right]
\end{equation}
as the diagonal matrix containing the photonic mode energies.
$H_{Landau}$ is the energy of the collective excitation electronic excitations.
The vector ${\bm \Omega}_{q_x}$ and the matrix ${\bm D}_{q_x}$ contain the coupling constants corresponding to $H_{int}$ and $H_{dia}$ respectively. ${\bm \Omega}_{q_x}$ is the collective vacuum Rabi frequency and comes from the resonant coupling between the cavity photons and the 2DEG. 
$H_{dia}$ (diamagnetic term by analogy with atomic physics) comes from the squared vector potential ${\bm A}^{2}_{em}$. In Appendix \ref{appHamiltonian}, we present the detailed derivation of these coupling constants. The final result is 

\begin{eqnarray}
\label{cstes}
{\bm \Omega}_{q_x} &=& \sqrt{\frac{2 \pi e^{2} \omega_{0}\, n_{QW} \rho_{2DEG}}{\epsilon\, m^{*} L_z}} \; \bar{\bm \omega}^{-\frac{1}{2}}_{q_x}\nonumber, \\ \\
{\bm D}_{q_x}&=& \frac{{\bm \Omega}_{q_x} \; {\bm \Omega}^{T}_{q_x}}{\omega_0}. \nonumber
\end{eqnarray}

with

\begin{equation}
\label{omegainv}
\bar{\bm \omega}^{-\frac{1}{2}}_{q_x} = diag\left[\omega^{-\frac{1}{2}}_{q_x,1},0,-\omega_{q_x,3}^{-\frac{1}{2}},0,\omega_{q_x,5}^{-\frac{1}{2}},0,-\omega_{q_x,7}^{-\frac{1}{2}}
,\dots \right]^T,   
\end{equation}

and the zeros at the even positions in Eq. (\ref{omegainv}) are given by the fact that, as apparent in Eq.
(\ref{shape}), only odd photonic modes are coupled to the electron gas.

If we consider only the resonant photonic mode, with frequency $\omega_{\bar{q}_x,\bar{n}_z} = \omega_0$, then
the corresponding normalized vacuum Rabi frequency reads
\begin{equation}
\frac{\Omega_{res}}{\omega_{0}} = \sqrt{\frac{2\alpha \, n_{QW} \nu}{\pi \sqrt{\epsilon}}}. 
\end{equation}
Apart from a geometric form factor of the order of $1$, we see that $\frac{\Omega_{res}}{\omega_{0}}  \propto \sqrt{\alpha \, n_{QW} \nu}$ as in Eq.  (\ref{ratio}), obtained with a back-of-the-envelope scaling calculation.

From Eq.(\ref{cstes}) it is clear that, for $\frac{\Omega_{res}}{\omega_0} \gg 1$, we have $D_{res} \gg \Omega_{res}$. In the ultrastrong coupling regime, we conclude that the ${\bm A}^2$ term becomes thus dominant over the vacuum Rabi coupling term.

\begin{figure}[h!]
\begin{center}
\includegraphics[width=270pt]{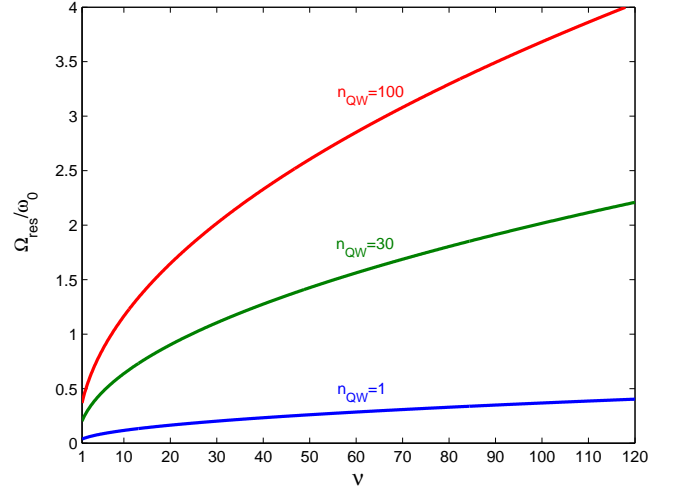}
\caption{\it The dimensionless vacuum Rabi coupling $\Omega_{res}/\omega_{0}$ versus the filling factor $\nu$ for different values of the number $n_{QW}$ of quantum wells. We consider only the photonic mode $\{q_x=0,n_z=1\}$ at resonance with the transition. Precisely, the resonance is defined such that $\omega_{q_x=0,n_z=1} = \omega_0$. Other parameters are given in the text.}
\label{rabi}  
\end{center}
\end{figure}

Because this Hamiltonian is quadratic in terms of ${\bm a}_{q_x}$ and $b_{q_x}$ operators, it can be exactly diagonalized using a generalized Hopfield transformation \cite{Gerace07}. We introduce the normal modes $p^{(i)}_{q_x}$ and $p^{\dagger (i)}_{q_x}$ (magnetopolaritons) defined as 

\begin{equation*}
\label{pop}
p^{(i)}_{q_x} = {\bm W}^{(i)}_{q_x} \; {\bm a}_{q_x} + X^{(i)}_{q_x} \; b_{q_x} +{\bm Y}^{(i)}_{q_x} \; {\bm a}_{-q_x}^{\dagger} + Z^{(i)}_{q_x} \; b_{-q_x}^{\dagger},
\end{equation*}

Given the bosonicity of ${\bm a}_{q_x}$ and $b_{q_x}$ operators, also the magnetopolariton operators satisfy the Bose commutation rule $\left[p^{(i)}_{q_x}, p^{\dagger (i)}_{q'_x} \right] = \delta_{i,i'} \delta_{q_x,q'_x}$ where the index $i$ runs over all the polariton branches. The condition for the total Hamiltonian to be diagonal in terms of magnetopolariton operators is 

\begin{equation}
\left[p^{(i)}_{q_x},H\right]= \hbar \omega^{(i)}_{q_x}\, p^{(i)}_{q_x}
\end{equation}

and the eigenvalue problem takes the form 

\begin{equation}
\mathcal{M}_{q_x} {\bm v}^{(i)}_{q_x} = \hbar \omega^{(i)}_{q_x} {\bm v}^{(i)}_{q_x}
\end{equation}

where $\hbar \omega^{(i)}_{q_x}$ correspond to the magnetopolariton energies, ${\bm v}^{(i)}_{q_x}$ is the vector $\left({\bm W}^{(i)}_{q_x}, X^{(i)}_{q_x}, {\bm Y}^{(i)}_{q_x},Z^{(i)}_{q_x}\right)^{T}$ and $\mathcal{M}_{q_x}$ is the infinite Hopfield matrix that is given by

\begin{equation}
\label{Matrix}
\mathcal{M}_{q_x} =
\left(\begin{array}{cccc}
{\bm \omega}_{q_x} +2 {\bm D}_{q_x} & i {\bm \Omega}_{q_x} & -2 {\bm D}_{q_x} & i {\bm \Omega}_{q_x}\\ \\ -i {\bm \Omega}^{T}_{q_x} & \omega_{0} & i {\bm \Omega}^{T}_{q_x} & 0 \\ \\
2 {\bm D}_{q_x} &  i {\bm \Omega}_{q_x} & -{\bm \omega}_{q_x} -2 {\bm D}_{q_x} & i {\bm \Omega}_{q_x} \\ \\ i {\bm \Omega}^{T}_{q_x} & 0 &- i {\bm \Omega}^{T}_{q_x} & -\omega_{0} \end{array} \right).
\end{equation} 

\section{Results}
\label{Results}

By diagonalizing the matrix in Eq. (\ref{Matrix}), we are now able to calculate the magnetopolariton energy dispersions in the multimode case.
For sake of simplicity, we have taken the lower photonic mode to be resonant with the cyclotron transition. 
In Fig. \ref{rabi}, we present results of the dimensionless vacuum Rabi frequency $\Omega_{res} /\omega_0$ as a function of the filling factor $\nu$ for different values of the number $n_{QW}$ of quantum wells. 
It is important to point out that the dimensionless coupling depends only on $\sqrt{\alpha \,n_{QW} \nu}$. Hence, different values of the two dimensional electron gas density $\rho_{2DEG}$, magnetic field $B$, and semiconductor effective mass $m^*$, can give rise to the same normalized vacuum Rabi coupling $\Omega_{res}/\omega_0$, provided that the filling factor $\nu$ stays constant. For example, a filling factor of $\nu = 100$ with a number of QWs of $n_{QW} = 100$ gives a dimensionless vacuum Rabi coupling that takes the impressive value $\Omega_{res} /\omega_0 \simeq 3.6$. This kind of parameters can be obtained using GaAs high-mobility samples with cyclotron frequencies in the microwave range, as already discussed at the end of Sec. \ref{system}. 

\begin{figure}[h!]
\begin{center}
\includegraphics[width=270pt]{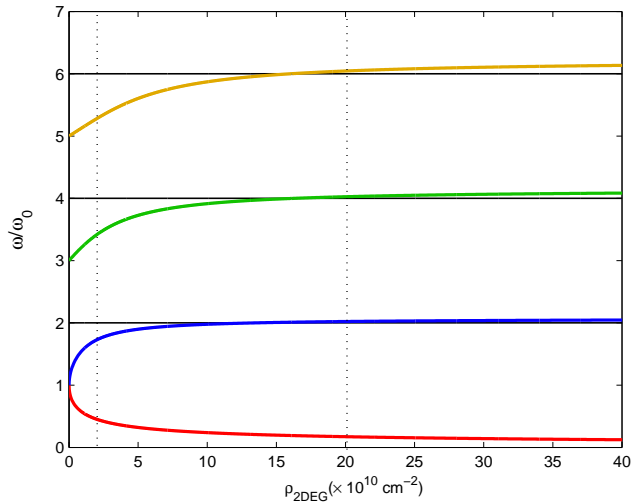}
\caption{\it \label{split} (Color online). Thick solid colored lines: normalized frequency of the first four magnetopolariton branches as a function of the 2D electron gas density $\rho_{2DEG}$ in the case where the energy of the first cavity mode $\{q_x=0,n_z=1\}$ is resonant to the cyclotron transition. Thin black solid lines depict the energies of the photonic branches, which are not coupled to the cyclotron transition (for the selection rules, see description in the text, in particular Eqs. (\ref{cstes}) and (\ref{omegainv})). The two vertical straight lines indicate the densities $\rho_{2DEG}=2 \; 10^{10} {\rm cm^{-2}}$ and $\rho_{2DEG}=2 \; 10^{11} {\rm cm^{-2}}$ (giving a  dimensionless vacuum Rabi coupling $\Omega_{res}/\omega_{0} \simeq 0.8$ and $\Omega_{res}/\omega_{0} \simeq 2.5$ respectively):  the corresponding frequency dispersions as a function of the wavevector $q_x$ are plotted in Figs. \ref{disp1} and \ref{disp2}. Parameters : $m^*=0.068 m_0$ (GaAs effective mass);  $B=40{\rm mT}$ (giving a cyclotron frequency $\omega_{0}=100{\rm GHz/rad}$); $n_{QW}=50$ and $L_{z}=0.25\, {\rm cm}$ which is chosen in such a way that $\omega_{q_x=0,n_z=1} = \omega_0$. With these parameters,  a good numerical convergence is obtained for a cutoff number of photonic  modes $\Lambda > 7$.} 
\end{center}
\end{figure}

\begin{figure}[h!]
\begin{center}
\includegraphics[width=270pt]{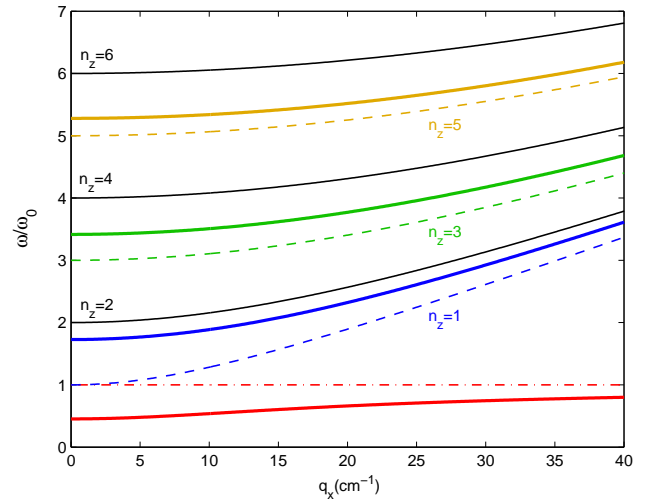}
\caption{\it \label{disp1} (Color online). Thick solid colored lines: normalized frequencies of the first four normal mode branches as a function of the wavevector $q_x$ for $\rho_{2DEG}=2 \; 10^{10} {\rm cm^{-2}}$ and $B=40{\rm mT}$ which correspond to a filling factor $\nu \simeq10$. The dashed color lines represent the dispersions of the photonic modes with odd values of $n_z$ that are coupled to the cyclotron transition. The red dashed-dotted line depicts the bare cyclotron frequency; the thin black solid lines represent the photonic modes with even values of $n_z$, which are not coupled to the electronic system. Other parameters are the same as in Fig. \ref{split}.}
\end{center}
\end{figure}

\begin{figure}[h!]
\begin{center}
\includegraphics[width=270pt]{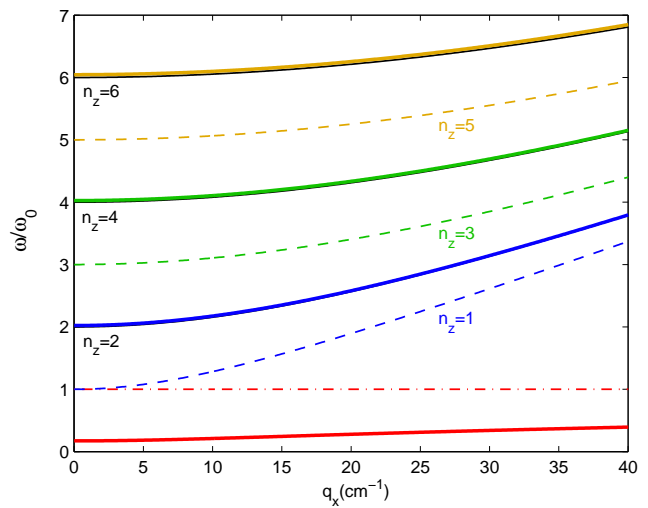}
\caption{\it \label{disp2} Same as Fig. \ref{disp1}, but with an electron density $\rho_{2DEG}=2 \; 10^{11} {\rm cm^{-2}}$ (corresponding to a filling factor $\nu \simeq 100$). }
\end{center}
\end{figure}

In Figure \ref{split}, the thick solid lines represent the normalized energies of the first four magnetopolariton branches as a function of the two dimensional electron gas density $\rho_{2DEG}$ in the case where the energy of the first cavity mode $\{q_x=0,n_z=1\}$ is equal to the cyclotron transition energy. 
Note that the thin solid lines depict the energies of the photonic modes that, due to the considered geometry, are not coupled to the cyclotron transition.
Fig. \ref{disp1} and \ref{disp2} depict a typical dispersion of the normalized frequencies of the first four polariton branches as a function of the wavevector $q_x$ for densities $\rho_{2DEG}=2 \; 10^{10} {\rm cm^{-2}}$ and $\rho_{2DEG}=2 \; 10^{11} {\rm cm^{-2}}$ respectively.

It is worth pointing out that for very large couplings the frequency of the lower branch asymptotically tends to $0$. This suggests that this kind of cavity excitations could affect significantly the magnetotransport properties of a cavity embedded two dimensional electron gas in the low frequency regime. 

\section{Conclusions and perspectives}
\label{Conclusions}
In conclusion, we have presented a quantum model describing the ultrastrong coupling between a cavity resonator and the cyclotron transition of a two dimensional electron gas. The present approach holds for the case of integer filling factors.
We have determined the second quantization light-matter Hamiltonian in terms of the electronic fermionic operators and of the cavity photon operators. We have derived and diagonalized an effective bosonic Hopfield-like Hamiltonian describing the magnetopolariton excitations. 
The ultrastrong coupling regime characterized by a vacuum Rabi frequency  much larger than the cyclotron transition frequency can be achieved with high filling factors, which are compatible with state-of-the-art GaAs high mobility two dimensional electron gas and cyclotron transitions in the microwave range. 
In our present approach, we have not considered the case of non-integer filling factors:  the nature of the magnetopolariton excitations may be qualitatively and quantitatively affected by a partially filled Landau band, an issue that needs to be explored in the future.

The role of Coulomb interaction between carriers and basic approximations have been critically discussed in Sec. (\ref{coulomb}).  
 
In the future, it will be interesting to explore the impact of the ultrastrong coupling on the magnetotransport of properties of a cavity embedded two dimensional electron gas. In fact, the energy of the lower magnetopolariton branch can become much smaller than $\hbar \omega_0$ for $\Omega_{res}/\omega_0 \gg 1$, meaning that even the low-frequency transport can be affected by the coupling to the cavity vacuum field. In particular, the Shubnikov de Haas oscillations, the nonlinear response to applied microwave field\cite{Mani14,Zudov16,Girvin}, could be strongly influenced by the presence of the cavity.

We would like to thank S. Barbieri, Y. Gallais, C. Sirtori and Y. Todorov for discussions.

\appendix

\section{Spatial shape of the cavity modes}
\label{modes}

The general form for the electromagnetic vector potential is 

\begin{equation}
\label{vector}
{\bm A_{em}}({\bm r}) =\sum_{\bm{q},\eta} \sqrt{\frac{2\pi\hbar c^{2}}{\epsilon \, \omega_{\bm{q}}}} \left(a_{\bm{q},\eta} {\bm u}_{\bm{q},\eta}+ a^{\dagger}_{\bm{q},\eta} {\bm u}^{*}_{\bm{q},\eta} \right),
\end{equation}
where the operator $a_{\bm{q},\eta}$ is the bosonic annihilation operator for a photon with polarization $\eta \in \{1,2 \}$ belonging to the mode labeled by the wavevector :
\begin{equation}
\label{mode2}
{\bm q} = \left ( \begin{array}{c}   q_{x} \\ q_{y} \\ q_{z} \end{array}  \right ) = \left ( \begin{array}{c}   q_x\\ \frac{\pi n_{y}}{L_{y}} \\ \frac{\pi n_{z}}{L_{z}} \end{array}  \right ) =  q \left ( \begin{array}{c}   \sin \theta \cos \phi \\ \sin \theta \sin \phi \\ \cos \theta \end{array}  \right ) ,
\end{equation}
where $q_x$ can vary continuously, while $q_y$ and $q_z$ are quantized ($n_y$ and $n_z$ are integer values). 
The spatial shape ${\bm u}_{\bm{q},\eta}$ of the modes is given by \cite{Kakazu12}  
\begin{eqnarray}
\label{mode}
{\bm u}_{{\bm q},1} & = & N e^{i q_{x} x} {\left(
\begin{array}{c}
i \sin (q_{z} z) \sin (q_{y} y) \, \cos \theta \cos \phi \nonumber \\   \nonumber \\ \sin (q_{z} z) \cos (q_{y} y) \, \cos \theta \sin \phi   \nonumber\\ \\ - \cos(q_{z} z) \sin (q_{y} y) \, \sin \theta \end{array} \right)},   \\ \\\nonumber  \\ 
{\bm u}_{{\bm q},2} & = &  N e^{i q_{x} x} {\left( 
\begin{array}{c}
- i \sin (q_{z} z) \sin (q_{y} y) \, \sin \phi  \\ \\ \sin (q_{z} z) \cos (q_{y} y) \, \cos \phi  \\\ \\ 0 \end{array} \right)}. \nonumber \ 
\end{eqnarray} 

The normalization constant $N$ reads 

\begin{equation}
N =
\begin{cases}
\sqrt{\frac{2}{V}} \; \; \text{if} \;\; q_y=0 \;\; \text{or} \;\; q_z =0 \;\\ \\
\frac{2}{\sqrt{V}} \; \; \text{otherwise}.
\end{cases}
\end{equation}

As it has been written in Sec. \ref{subA} and \ref{subB}, the condition $L_y<<L_z$ allows us to neglect all the modes with $q_y \neq 0$. From Eq. (\ref{mode2}), we see that $q_y=0$ implies $\phi =0$ and consequently only the polarization $\eta =2$ is present. Omiting the $q_y=0$ and $\eta=2$ indexes, we can see from Eqs. (\ref{vector}), (\ref{mode2}) and (\ref{mode}) that the electromagnetic vector potential and the spatial shape of the cavity modes are given by Eq. (\ref{vector1}) and (\ref{shape}) respectively.

\section{Landau Levels in the Landau gauge}
\label{appLandau}

The one electron wavefunctions in the Landau gauge (${\bm A}_{0}=-By\, {\bm u}_{x}$), reads 

\begin{equation}
\varphi_{n,k,m} ({\bm r}) = \chi_{k} (x) \; \phi_{n} (y-y_{0}) \; \xi_{m} (z),
\end{equation}

where

\begin{eqnarray*}
\chi_{k} (x) &=& \frac{1}{\sqrt{L_{x}}} e^{i k x}, \\ 
\phi_{n} (y-y_{0})&=& \frac{1}{\sqrt{2^{n} n! l_{0} \sqrt{\pi}}} \, H_{n} \left(\frac{y-y_{0}}{l_{0}}\right) e^{-\frac{\left(y-y_{0}\right)^{2}}{2l_{0}^2}}, 
\end{eqnarray*}

$y_0 =k l^{2}_{0}$ is the so-called guiding center position depending on $k$, $\xi_{m} (z)$ is the confinement wavefunction of the $m$-th conduction subband of the quantum well and $H_n$ is the Hermite polynomial of degree $n$. With the chosen gauge the wavefunction is thus factorized along the three axis in a plane wave, an harmonic oscillator wavefunction and a confinement-dependent function.
For an infinitely deep quantum well with width $L_{QW}$, we can find an analytic form for the confinement-dependent function

\begin{equation}
\xi_{m} (z) =
\begin{cases}
\sqrt{\frac{2}{L_{QW}}} \cos\left[\frac{m \pi \left(z-\frac{L_{z}}{2}\right)}{L_{QW}}\right] \; \; m \;\text{odd}\\ \\
\sqrt{\frac{2}{L_{QW}}} \sin\left[\frac{m \pi \left(z-\frac{L_{z}}{2}\right)}{L_{QW}}\right] \; \; m \;\text{even}.
\end{cases}
\end{equation}

As explained in section \ref{subC}, we omit both the spin quantum number, the Zeeman splitting of the LLs and the $m$ index that do not play any role here.

Knowing that $k=2 \pi n_{x} /L_x$ with $n_x \in \mathbb{N}$, the many body electronic ground state reads

\begin{equation}
\ket{F}=\prod_{j=1}^{n_{QW}} \prod_{n=0}^{\nu-1} \prod_{n_{k}=1}^{\mathcal{N}} c_{n,k}^{(j)\dagger} \ket{0},
\end{equation}
where $\ket{0}$ is the empty conduction band state and $c_{n,k}^{(j)\dagger}$ is the fermionic operator creating an electron in the $n^{th}$ Landau level with wavevector $k$ along $x$ in the $j^{th}$ quantum well.

\section{Second quantization hamiltonian }
\label{appHamiltonian}
The minimal coupling Hamiltonian describing the electrons coupled to the electromagnetic field reads:

\begin{equation}
\label{Hmin}
H=\sum_{i} \frac{1}{2m^{*}} \left({\bm p}_{i}-\frac{e}{c}{\bm A}({\bm r}_{i})\right)^2 + H_{cavity},
\end{equation}

where the sum runs over all the electrons (in all the quantum wells) and $H_{cavity}$ describes the free electromagnetic field.
Developing the square in Eq. (\ref{Hmin}) and writing ${\bm A} ({\bm r}) = {\bm A_{0}} ({\bm r}) + {\bm A_{em}} ({\bm r})$ (as in Eq. (\ref{expressionA})), we can identify four different terms in the Hamiltonian, namely

\begin{equation} 
H = H_{Landau} + H_{int} + H_{dia} + H_{cavity}.
\end{equation}

$H_{Landau} = \sum_{j} H_{Landau}^{(j)}$ and $H_{int}= \sum_{j} H_{int}^{(j)}$. $H_{cavity}$ describes the free quantum cavity electromagnetic field and has been expressed in terms of second quantized photon operators in Eq. (\ref{field}).
$H_{dia}$ comes from the ${\bm A}^2$ term.
$H^{(j)}_{Landau}$ is the Hamiltonian describing the electrons in the $j^{th}$ quantum well, interacting with the static magnetic field, giving rise to the Landau levels. Finally $H^{(j)}_{int}$ describes the interaction of the electrons in the $j^{th}$ quantum well with the cavity electromagnetic field. 
In the following we will express $H^{(j)}_{Landau}$, $H^{(j)}_{int}$ and $H_{dia}$ in terms of the ${\bm a}_{q_x}$ and $c^{(j)}_{n,k}$ operators introduced in section \ref{subB}, \ref{subC} and Appendix \ref{appLandau}.

$H^{(j)}_{Landau}$ is by construction a diagonal operator, because its eigenmodes are chosen as a basis for the transition in second quantization formalism, namely:
\begin{equation}
\label{LN}
H_{Landau}^{(j)} = \sum_{n,k} n \hbar \omega_{0}\, c_{n,k}^{(j)\dagger} c_{n,k}^{(j)},
\end{equation}

As already mentioned in Sec. \ref{subC}, because the electronic excitations are collective, we have to express $H^{(j)}_{Landau}$ in the subspace of the collective excitations.
By calculating the energy of such collective excitations (i.e. the matrix element $\bra{F} b_{q_x} H_{Landau} \, b_{q_x'}^{\dagger} \ket{F}$, where $\ket{F}$ is  the electronic Fermi ground state defined in Appendix \ref{appLandau}), it is straightforward to find in such subspace
$H_{Landau}$ can be replaced by  $\sum_{q_{x}} \hbar \omega_{0} \, b_{q_x}^{\dagger}b_{q_x}$, where we have omitted a constant energy term.

In order to express $H_{int}$ in terms of second quantized operators, we have to calculate the matrix elements of the form

\begin{widetext}
\begin{equation}
\label{matel}
\bra{n,k,m} {\bm p} \cdot {\bm A}_{em} ({\bm r})\ket{n',k',m'} \propto \sum_{q_{x},n_z}\, \bra{m}\sin(\pi n_z z/L_{z})\ket{m'} \, \bra{n} p_{y} \ket{n'}  \left(a_{q_{x},n_z} \bra{k} e^{i q_{x} x} \ket{k'} + a_{q_{x},n_z}^{\dagger} \bra{k} e^{-i q_{x} x} \ket{k'}\right),
\end{equation}
\end{widetext}

where $\ket{n,k,m}$ is the state of an electron in the $m^{th}$ subband and $n^{th}$ Landau level with momentum $k$ (see appendix \ref{appLandau}). As already mentioned if we assume that $L_{QW}\ll L_{z}$, it comes from the first term in the right hand side of Eq. (\ref{matel}) that $m=m'$.

The matrix elements for the transition between Landau levels read 

\begin{eqnarray*}
\bra{n} p_{y} \ket{n'} = \frac{i \, m^{*} \omega_{0} \, l_{0}}{\sqrt{2}\,\left(n-n' \right)} \left(\sqrt{n'+1} \, \delta_{n,n'+1} + \sqrt{n'} \, \delta_{n,n'-1}\right).
\end{eqnarray*}

Therefore, we get the harmonic oscillator selection rule $n'=n\pm1$;  transitions such that $\vert n'-n \vert \geq 2$ are strictly forbidden.
 
The integration over the $x$ direction gives rise to the momentum conservation and we obtain the result

\begin{eqnarray*}
 H_{int}^{(j)} &=& \sum_{q_x,k} i \hbar \;{\bm \chi}^{T}_{q_x} {\bm a}_{q_x} \left(c_{\nu,k+q_x}^{(j)\dagger} c_{\nu-1,k}^{(j)} - c_{\nu-1,k}^{(j)\dagger} c_{\nu,k-q_x}^{(j)} \right ) \\ \\
  &+& \sum_{q_x,k} i \hbar \;{\bm \chi}^{T}_{q_x} {\bm a}^{\dagger}_{q_x} \left(c_{\nu,k-q_x}^{(j)\dagger} c_{\nu-1,k}^{(j)} - c_{\nu-1,k}^{(j)\dagger} c_{\nu,k+q_x}^{(j)} \right ).
\end{eqnarray*}

The coupling constant (we are using the same notation as in section \ref{subC}) is given by

\begin{equation}
\label{chi}
{\bm \chi}_{q_x}=\sqrt{\frac{2 \pi e^{2} \omega_{0} \; \nu}{\epsilon \,m^{*} S L_z}} \; \bar{\bm \omega}^{-\frac{1}{2}}_{q_x},
\end{equation}

Introducing the bosonic bright mode creation operator as described in section \ref{subC}
\begin{equation}
b_{q_x}^{\dagger} = \sqrt{\frac{\nu}{n_{QW}\rho_{2DEG}S}}\; \sum_{j,k} c_{\nu,k+q_x}^{(j)\dagger} \, c_{\nu-1,k}^{(j)},
\end{equation} 

we can check that the interaction Hamiltonian $H_{int}$ can be exactly rewritten as it is given in Eq. (\ref{terms}) by taking
\begin{equation*}
\label{Omega}
{\bm \Omega}_{q_x} = {\bm \chi}_{q_x} \sqrt{ \frac{n_{QW} \rho_{2DEG} S} {\nu}}. 
\end{equation*}

Analogously in order to calculate $H_{dia}$ we need  the matrix elements : 

\begin{widetext}
\begin{equation}\begin{array}{rl}
\bra{n,k,m} {\bm A}_{em}^{2} \ket{n',k',m'} & \propto \sum_{q_{x},q_{x}',n_z,n'_z} \bra{m} \sin(\pi n_z z/L_{z}) \sin(\pi n'_z z/L_{z}) \ket{m'} \\ \\ 
& \times \left \{ a_{q_x,n_z}^{\dagger} \, a_{q'_x,n'_z} \bra{n,k} e^{i \left(q'_x - q_x \right) x} \ket{n',k'} + a_{q_x,n_z} \, a_{q'_x,n'_z}^{\dagger} \bra{n,k} e^{-i \left(q'_x - q_x \right) x} \ket{n',k'} \right . \\ \\ 
& \left .+ a_{q_x,l_z}^{\dagger} \, a_{q'_x,l'_z}^{\dagger} \bra{n,k} e^{-i \left(q'_x + q_x \right) x} \ket{n',k'} + a_{q_x,l_z} \, a_{q'_x,l'_z} \bra{n,k} e^{i \left(q'_x + q_x \right) x} \ket{n',k'}
\right \}.
\end{array}
\end{equation}
\end{widetext}
Given that the quantum well size $L_{QW} \ll L_z$, the integral along $z$ gives the same selection rule as for $H_{int}$ : $m=m'$ so $\bra{m} \sin(\pi l_z z/L_{z}) \sin(\pi l'_z z/L_{z}) \ket{m'} \approx \sin(\pi n_z/2) \sin(\pi n'_z/2) \delta_{m,m'}$. 

On the other hand, the matrix elements along the other directions $x$ and $y$ can be factorized. For example the term $\bra{n,k} e^{i \left(q'_x - q_x \right) x} \ket{n',k'}$ can be factorized as

\begin{equation}
\label{factor}
\bra{k} e^{i \left(q'_x - q_x \right) x} \ket{k'} \langle n,k \vert n',k' \rangle.
\end{equation}

The second term of the previous equation is then given by the overlap integral $I_{n,n'} \left(k,k'\right)$  

\begin{equation}
\label{guiding}
I_{n,n'} \left(k,k'\right)=\int_{0}^{L_{y}} \! dy \; \phi_{n} \left(y-y_{0} (k)\right) \phi_{n'} \left(y-y_{0} (k')\right).
\end{equation}

Assuming $L_{y}>>l_{0}$ (which is realistic being $l_{0} \approx 0.1{\rm \mu m}$ for $\omega_{0}=52 {\rm GHz \, rad^{-1}}$) we get :

\begin{equation}
\label{guiding1}
I_{n,n'} \left(k,k'\right) \approx \int_{-\infty}^{+\infty} \! dy \; \phi_{n} \left(y \right) \phi_{n'} \left(y-\Delta y_{0} \right).
\end{equation}

with $\Delta y_{0} = y_{0} (k') -y_{0} (k)$. We can then use the useful expansion of the harmonic oscillator wavefunctions \cite{Smith92} 
 
\begin{equation}
\label{guiding2}
\phi_{n} \left(y-\Delta y_{0} \right) =  \sum_{m=0} \frac{n!}{m!} \left(\beta^{\frac{1}{2}} \right)^{m-n} e^{-\beta/2}\; L^{m-n}_{n} \left(\beta \right) \phi_{m} \left(y \right).
\end{equation}

where $\beta = \frac{\Delta y_{0}^{2}}{2l^{2}_{0}}$ and $L^{m-n}_{n}$ is the Laguerre polynomial of degree $n$ and index $m-n$. Substituting Eq. (\ref{guiding2}) into Eq. (\ref{guiding1}), we obtain 

\begin{equation}
\label{guiding3}
I_{n,n'} \left(k,k'\right) \approx \frac{n'!}{n!} \left(\beta^{\frac{1}{2}} \right)^{n-n'} e^{-\beta/2}\; L^{n-n'}_{n'} \left(\beta \right).
\end{equation}

Because of the gaussian dependence $\propto e^{-\frac{\Delta y_{0}^{2}}{4l^{2}_{0}}}$, $I_{n,n'} \left(k,k'\right)$ exhibits a peak centered at $\Delta y_{0} = 0$ as sharp as $l_{0}$. Thanks to the definition $y_{0} (k) =l^{2}_{0} \, k$ and the condition that $l_{0}<<L_{y}$, we can consider approximatively that $I_{n,n'} \left(k,k'\right) \propto \delta_{k,k'}$. In addition, due to the orthonormality of the function $\phi_n$, setting $k=k'$ (or equivalently $\Delta y_{0} = 0$) into the integral (\ref{guiding1}) implies that $I_{n,n'} \left(k,k'\right) \approx \delta_{n,n'}$.
  
Hence, setting $k=k'$ into the integral given by the first term of eq. (\ref{factor}), we obtain the corresponding selection rules on $q_x$ and $q'_x$. Note that in the case of the previous example, we get that $q'_x=q_x$. Finally, we find that $H_{dia}$ is given by Eq. (\ref{terms}) with the corresponding coupling constants.


\begin{thebibliography}{99}

\bibitem{Thompson92}  R. J. Thompson, G. Rempe and H. J. Kimble, Phys. Rev. Lett. {\bf 68}, 1132 (1992). 

\bibitem{Raymond01}
J. M. Raimond, M. Brune and S. Haroche, Rev. Mod. Phys. {\bf 73}, 565 (2001).

\bibitem{Weisbuch92} C. Weisbuch, M. Nishioka, A. Ishikawa and Y. Arakawa, Phys. Rev. Lett. {\bf 69}, 3314 (1992).

\bibitem{Dini03}
D. Dini, R. K\"ohler, A. Tredicucci, G. Biasiol and L. Sorba, Phys. Rev. Lett. {\bf 90}, 116401 (2003).

\bibitem{Wallraff04}
A. Wallraff, D. I. Schuster, A. Blais, L. Frunzio, R.-S. Huang, J. Majer, S. Kumar, S. M. Girvin and R. J. Schoelkopf, Nature {\bf 431}, 162 (2004).

\bibitem{Ciuti05}
C. Ciuti, G. Bastard and I. Carusotto, Phys. Rev. B {\bf72}, 115303 (2005).

\bibitem{Ciuti06}
C. Ciuti and I. Carusotto, Phys. Rev. A {\bf 74}, 033811 (2006).

\bibitem{Devoret07}  M. Devoret, S. Girvin and R. Schoelkopf,  Ann. Phys. (Leipzig) {\bf 16}, 767  (2007).

\bibitem{Anappara09}
Aji A. Anappara, Simone De Liberato, Alessandro Tredicucci, Cristiano Ciuti, Giorgio Biasiol, Lucia Sorba, and Fabio Beltram, Phys. Rev. B {\bf 79}, 201303(R) (2009). 

\bibitem{Yanko09} Y. Todorov, A. M. Andrews, I. Sagnes, R. Colombelli, P. Klang, G. Strasser, and C. Sirtori, Phys. Rev. Lett. {\bf 102}, 186402 (2009).

\bibitem{Yanko10} Y. Todorov {\it et al.}, unpublished results.  

\bibitem{Bourassa} J. Bourassa, J. M. Gambetta, A. A. Abdumalikov, Jr., O. Astafiev, Y. Nakamura, and A. Blais, Phys. Rev. A {\bf 80}, 032109 (2009).

\bibitem{Nataf} P. Nataf, C. Ciuti, Phys. Rev. Lett. {\bf 104}, 023601 (2010).

\bibitem{Kardar99}
M. Kardar and R. Golestanian, Rev. Mod. Phys. {\bf 71}, 1233 (1999).

\bibitem{Gunter09}
G. G\"{u}nter, A. A. Anappara, J. Hees, A. Sell, G. Biasiol, L. Sorba, S. De Liberato, C. Ciuti, A. Tredicucci, A. Leitenstorfer and R. Huber, Nature {\bfÊ458}, 178-181 (2009). 

\bibitem{DeLiberato07} 
S. De Liberato, C. Ciuti and I. Carusotto, Phys. Rev. Lett. {\bf 98}, 103602 (2007).

\bibitem{DeLiberato09}
S. De Liberato, D. Gerace, I. Carusotto and C. Ciuti, Phys. Rev. A {\bf 80}, 053810 (2009).

\bibitem{Dicke54}
R. H. Dicke, Phys. Rev {\bf 93}, 99 (1954).

\bibitem{Ritchie} A. V. Polisskii, V. I. Talyanskii, J. Cole, C. G. Smith, M. Pepper, D. A. Ritchie, J. E. F. Frost, and G. A. C. Jones
J. Appl. Phys. {\bf 72} 4736 (1992). 


\bibitem{Hopfield} J. J. Hopfield, Phys. Rev. {\bf 112}, 1555 (1958).

\bibitem{Savona94}
V. Savona, Z. Hradil, A. Quattropani and P. Schwendimann,
Phys. Rev. B {\bf 49}, 8774 (1994).

\bibitem{Gerace07}
D. Gerace and L. C. Andreani, Phys. Rev. B {\bf 75}, 235325 (2007).


\bibitem{Carusotto08}
I. Carusotto, M. Antezza, F. Bariani, S. De Liberato, and C. Ciuti,
Phys. Rev. A, {\bf 77}, 063621 (2008).



\bibitem{Aleiner} I. L. Aleiner and L. I. Glazman, Phys. Rev. B {\bf 52}, 11296 (1995).

\bibitem{Kohn13} W.Kohn, Phys. Rev. {\bf 123}, 1242 (1961).

\bibitem{Mani14} 
R. G. Mani, J. H. Smet, K. von Klitzing, V. Narayanamurti, W. B. Johnson and V. Umansky, Nature {\bf 420}, 646 (2002).

\bibitem{Zudov16} 
M. A. Zudov, R. R. Du, L. N. Pfeiffer, and K. W. West, Phys. Rev. Lett. {\bf 90}, 046807 (2003).

\bibitem{Girvin}
A. C. Durst, Subir Sachdev, N. Read and S. M. Girvin, Phys. Rev. Lett. {\bf 91}, 086803 (2003).

\bibitem{Kakazu12}
K. Kakazu and Y. S. Kim, Phys. Rev. A {\bf 50}, , 1830 (1994).

\bibitem{Smith92}
W. L. Smith, J. Mol. Spec {\bf 225}, 39 (2004).

\end{thebibliography}
\end{document}